\documentclass[aps,pra,a4paper,twocolumn,10pt,amsmath,amssymb,superscriptaddress]{revtex4-1}

\usepackage{graphicx,eurosym}

\newcommand{\ket}[1]{|#1\rangle}

\begin{document}

\title{From Quantum Optics to Quantum Technologies}
\author{Dan Browne}
\affiliation{Department of Physics and Astronomy, University College London, Gower Street, London WC1E 6BT, UK}
\author{Sougato Bose}
\affiliation{Department of Physics and Astronomy, University College London, Gower Street, London WC1E 6BT, UK}
\author{Florian Mintert}
\affiliation{QOLS, Blackett Laboratory, Imperial College London, SW7 2AZ, UK}
\author{M. S. Kim}
\affiliation{QOLS, Blackett Laboratory, Imperial College London, SW7 2AZ, UK}

\begin{abstract}
Quantum optics is the study of the intrinsically quantum properties of light. During the second part of the 20th century experimental and theoretical progress developed together; nowadays quantum optics provides a testbed of many fundamental aspects of quantum mechanics such as coherence and quantum entanglement.
Quantum optics helped trigger, both directly and indirectly, the birth of quantum technologies, whose aim is to harness non-classical quantum effects in applications from
 quantum key distribution to quantum computing. Quantum light remains at the heart of many of the most promising and potentially transformative quantum technologies.
In this review, we celebrate the work of Sir Peter Knight and present an overview of the development of quantum optics and its impact on quantum technologies research. We describe the core theoretical tools developed to express and study the quantum properties of light, the key experimental approaches used to control, manipulate and measure such properties and their application in quantum simulation, and quantum computing.
  
\end{abstract}

\maketitle

\section{Introduction}

In 1900, Max Planck postulated that the energy of the light field was quantised, triggering the birth of quantum mechanics which become one of the central pillars of modern physics.
The development of the laser in 1960 provided a precise new tool for the generation of coherent monochromatic light, and the field of quantum optics flourished through close alignment of experimental and theoretical research.

In this review, we celebrate the works of Sir Peter Knight, who is known for many pioneering works, influential reviews and textbooks~\cite{Knight6,Knight7} in quantum optics. Quantum optics was instrumental in the development of quantum technology and quantum information and Peter Knight is one of the founding fathers in this development. 

Quantum optics provided the tools to study the foundations of quantum mechanics with exquisite precision. The interaction of an isolated atom and a light field provided the ideal  testbed for controversial ideas. For example, ``quantum jumps'' are a manifestation of the discontinuity inherent in the quantum measurement process,~\cite{Cook85,Knight3} which can be directly observed~\cite{Bergquist1}.  One of the key concepts of quantum optics is the coherent state, introduced by Schr\"{o}dinger, and used by Glauber and Sudarshan to study high-order coherences of light~\cite{Glauber63, Sudarshan63}. Squeezed light,~\cite{Knight1} in which the  quantum noise is reduced in one quadrature at the expense of the increased noise in the other quadrature, has also played a central role in quantum optics as it developed. 
From the beginning, quantum optics combined the study of fundamental physics with applications to technology. Squeezed light, for example, enables a new kind of  precision measurement, with application, for instance, in gravitational wave detection~\cite{Asasi13} and for noiseless communications~\cite{Yuen78}. 

The Jaynes-Cummings model (JCM)~\cite{JCM,Knight2}, which describes the interaction between a two-level atom and a single mode of the electromagnetic field,
 provided a lens for studying the nonclassical properties  of the atom-field interaction. One of the important consequences of the quantisation of the field was the collapses and revivals of Rabi oscillations~\cite{Eberly80} which was tested in a  cavity quantum electrodynamics (QED)   setup 
 ~\cite{Walther,Meekhof}.  An information-theoretic approach~\cite{Knight5} for the JCM found that the cavity field which was initially prepared in a large amplitude coherent state will become a coherent superposition state at a certain interaction time~\cite{Gea,KK}. 

While quantum optics focusses on the physics of light and atoms, quantum information focusses on the properties and applications of the qubit. A qubit, or quantum bit, 
is the quantum-mechanical extension of a conventional bit $\{0, 1\}$. As a quantum state, the state of a qubit $|\psi\rangle$ can be in a superposition of 0 and 1:
\begin{equation*}
|\psi\rangle=\alpha|0\rangle + \beta |1\rangle
\label{qubit}
\end{equation*}
where $\alpha$ and $\beta$ are complex numbers satisfying $|\alpha|^2+|\beta|^2=1$. The number of complex numbers needed to describe the state of a quantum system grows exponentially with its complexity (for example, the numbers of its constituents) which soon makes it prohibitively difficult to perform exact calculations of its behaviour, particularly those emergent properties which cannot be easily approximated or guessed by human ingenuity. This fact lead Feynman~\cite{Feynman} to suggest, in the early 1980s, that the properties of complex quantum systems should ideally be studied with collections of controllable and programmable quantum entities which interact with each other to ``mimic" the behaviour of the system being studied. 

The counterpart to coherence in quantum optics is the quantum mechanical superposition of the qubit. Extended across multiple qubits, superposition leads to entanglement, and information can be represented and processed in an intrinsically non-classical way. Thus a quantum computer~\cite{Deutsch} would be able to compute in an intrinsically different way to standard classical computers. Quantum computing became the subject of intense study after  Shor invented an efficient quantum algorithm for factoring, a problem for which no efficient classical algorithm is known despite centuries of study~\cite{Shor}. 

Another application of quantum mechanics in information processing is found in sending secret messages. It is well-known that  a one-time pad (a string of random numbers shared by two people) is the most secure way to send a secret message. In 1984, Bennett and Brassard~\cite{BB} invented a scheme to
create a one-time pad between distant partners using non-classical light fields prepared in superposition states (\ref{qubit}). 

The current development of quantum technology and quantum information processing is based on our ability to control coherences of a large quantum system. In doing this one of the main obstacles is the system's uncontrollable
interaction with  its environment.
There have been extensive studies of decoherence mechanisms in the framework of Markovian and non-Markovian open quantum systems and quantum trajectories. Knight worked on the sources and characteristics of quantum decoherence using a quantum jump approach~\cite{Knight4} with his colleague Martin Plenio. He also worked on various ways to protect a quantum state from decoherence using the phase control of the system~\cite{Knight8} and the concept of decoherence-free subspace~\cite{Knight9}. A key to overcome the decoherence in quantum computation came from quantum error correction which requires a large number of entangled resources~\cite{Shor-error}. The theory of quantum error correction was further developed with the concept of fault-tolerant quantum computation~\cite{Steane}.

Quantum entanglement is one of the most important non-classical ingredients in quantum technology. We say two quantum systems are entangled when the density operator of the total system cannot be written as a weighted sum of the product states:
\begin{equation}
\hat{\rho}\neq \sum_{n=0} P_n\hat{\rho}_{1,n} \otimes \hat{\rho}_{2,n}
\label{ent}
\end{equation}
where $P_n$ is a probability function and $\hat{\rho}_{i,n}$ is a density operator of system $i=1,2$.
Entanglement can be thought of as a special kind of correlation where it is possible for the states of local systems to carry more uncertainty than
the state of the global system, something impossible in classical statistics. Entanglement, which is closely connected to nonlocality in the Einstein-Podolsky-Rosen (EPR) paradox~\cite{EPR}, has had a number of direct applications. Ekert~\cite{Ekert} proposed a scheme to share a one-time pad between the two  users of a secret message, using the entangled nonlocal state of light fields. This  is an archetypical example to connect the very foundational issue of quantum mechanics to information technology. One important development in quantum information processing is quantum teleportation where entanglement of the quantum channel is a key for success of the protocol. A long-haul communication of quantum superposition states is highly challenging because such the quantum state is subject to decoherence. An alternative is teleporting the quantum state. Even though the entangled quantum channel is also subject to decoherence, it is in theory possible to distill entanglement to extract a small number of highly entangled pairs from a large number of weakly entangled pairs~\cite{Bennett96-1,Bennett96-2}. A simple but powerful distillation protocol~\cite{Knight10} was developed using entanglement swapping~\cite{Zukowski93,Knight11}.  Entanglement can be described by using the Schmidt decomposition~\cite{Knight12}. Finding a measure of entanglement was one of the issues to characterise quantum entanglement, Peter Knight and his colleagues pioneered to find useful quantification of quantum entanglement~\cite{Knight13,Vedral98,Lee00}. 

Atom optics, which is an effort to use quantum coherences of atomic motion, is another branch of quantum optics. Indeed, a very early form of atom interferometry is the Ramsey interferometry based on the quantum coherences in atomic internal states. Atomic clocks have been developed and precision sensors and accelerometers have been investigated based on atom optics. 

In this review, we survey how quantum optics has developed into quantum technology highlighting the roles and characteristics of photons, ions, atoms and mechanical oscillators. In Section~\ref{PHOTONsec}, we review the quantum properties of light itself, focussing on photons and coherent states. In Section~\ref{jcm}, we see how the JCM provides a powerful tool for the study of the interaction of quantum light and atoms. In Section~\ref{ion-traps}, we review trapped ions, an important test-bed for these theoretical ideas and a platform for the development of quantum technologies. In Section~\ref{opto}, we introduce mechanical oscillators, representing an analogue of quantum light in meso-scopic matter. Finally, in Section~\ref{QT_sec} we review how the application of these ideas and tools for the precise manipulation of quantum systems is giving rise to new technologies.

\section{Photons}\label{PHOTONsec}
The study of photons is at the heart of quantum optics. The concept of ``particles of light" has gone through various stages. While the quantisation of energy was suggested to explain the blackbody radiation at the initial stage of quantum mechanics, it was not clear which energy was really quantised as well as its consequences. 
Glauber~\cite{Glauber63} defined the coherent state of photons as an eigenstate of the annihilation operator and found that the coherent state is represented as a displaced vacuum in phase space. The coherent state is known to describe the photon number statistics and the coherence properties of a laser field. 

\subsection{Nonclassicalities}
Characterisation of nonclassical behaviour, which cannot be described by classical theories, was an important issue of quantum optics and photons were at the heart of this study. Antibunching and sub-Posssonian statistics were investigated. While a laser field was invented based on a quantum-mechanical interaction between atoms and light fields, the properties of the laser field could be understood based on the theory of stable light field in classical optics. As early as 1980, Knight and Milonni investigated a fully quantum-mechanical atom-field interaction model and signatures of nonclassical behaviours of light fields~\cite{Milonni}. 

The single-mode coherent field is represented by the Poissonian photon statistics and its intensity correlation $g^{(2)}(\tau)$ of the field is 1, where $g^{(2)}$ is defined as
\begin{equation}
g^{(2)}(\tau)=\frac{\langle{\cal T}:\hat{I}(t)\hat{I}(t+\tau):\rangle}{\langle\hat{I}(t)\rangle\langle\hat{I}(t+\tau)\rangle} 
\label{int-cor}
\end{equation}
with $\cal T$ and $:~:$ denoting time- and normal-ordering of the operators respectively and $\hat{I}$ is the intensity operator. Throughout the paper, an operator is represented by $\hat{}$. Using the Cauchy inequality, it can be shown that the intensity correlation function is larger than or equal to unity if the field is not quantised~\cite{Knight7}. Hong and Mandel~\cite{Hong86} experimentally demonstrated sub-Poissonian photon statistics of a single photon state which was generated, conditioned on a single photon measurement of a twin beam from a parametric conversion process.  As it is clear in the definition of the single photon state, once it is detected at any time, there is nothing left to be detected at any other time; hence sub-Poissonican photon statistics with $g^{(2)}=0$. Another consequence of Cauchy's inequality on the classical intensity correlation function is $g^{(2)}(\tau)<g^{(2)}(0)$: light fields are more strongly correlated without a time delay than with it. This so-called bunching effect is a characteristic of a classical field. Kimble, Dagenais and Mandel~\cite{Kimble77} experimentally demonstrated antibunching photon correlation in a field scattered by single sodium atoms excited by a dye laser. 

In mechanics, dynamics of a system is often studied in phase space composed of momentum and position - more generally, two canonically conjugate variables, $p$ and $q$. At a given time, a physical system is described by a joint probability function ${\cal P}(p,q)$. All the statistical properties of the system at that time can then be calculated using a simple probability theory:
$
\langle p^n q^m\rangle =\int p^n q^m {\cal P}(p,q)dpdq 
 = \langle q^m p^n\rangle.
$
Had there been a quantum joint probability ${\cal P}'$, we would have the following
\begin{eqnarray}
\langle \hat{p}^n\hat{q}^m\rangle = \int p^n q^m {\cal P}'(p,q)dpdq \nonumber = \langle \hat{q}^m \hat{p}^n\rangle.
\label{q-joint}
\end{eqnarray}
However, in quantum mechanics, the two canonically conjugate operators do not commute thus $\langle \hat{p}^n\hat{q}^m\rangle \neq \langle \hat{q}^m\hat{p}^n\rangle$ and Eq.(\ref{q-joint})  can not apply to quantum mechanics.
Hence it is clear that there is no joint probability in quantum-mechanical phase space. Noticing this, Wigner derived a probability-like function $W(p,q)$ which satisfies the marginality in probability theory; $\int W(p,q)dp={\cal P}(q)$ and $\int W(p,q)dq={\cal P}(p)$ where ${\cal P}(q)$ and ${\cal P}(p)$ are the marginal probability functions in $p$ and $q$ which are well-defined in quantum mechanics. It was later found that the Wigner function~\cite{Wigner} is the unique function which satisfies the marginality in quantum mechanics~\cite{Klauder}. Since then, a few quasiprobability functions have been suggested, notably by Husimi~\cite{Husimi}, Glauber~\cite{Glauber63-2} and Sudarshan~\cite{Sudarshan}, 

Even though the Wigner function shares the characteristics of the probability function, it is not a probability function. In particular, it can have negative values at some points of the phase space unlike a probability function and negativity of a Wigner function is known to be a nonclassical behaviour, which was demonstrated experimentally for a field having interacted with a single atom~\cite{Haroche}. Due to the over-completeness of the coherent state, any density operator can be represented as a sum of diagonal matrix elements in the coherent state basis. 

\subsection{Parametric downconversion}
One important nonclassical state in quantum optics is a squeezed state~\cite{Knight1} whose quadrature noise is reduced under the vacuum limit. In particular, a squeezed vacuum is generated by the spontaneous parametric down conversion (SPDC). As early as 1970, Burnham {\it et al.}~\cite{Burnham70} suggested to generate optical photon pairs using the $\chi^{(2)}$-nonlinear optical process. Since then, the SPDC has indeed been one of the most frequently used processes in quantum optics, not only to generate squeezed states but also, for example, to generate single photons~\cite{Hong86} and entangled photons. 
The SPDC converts a pump photon to two daughter photons whose frequency and propagation direction are determined by the principles of momentum and energy conservation, which leads the two daughter photons correlated in momentum and energy. Considering the probability of the number of daughter photons generated decreases exponentially, the quantum state of the daughter photons is represented by
\begin{equation}
|\Psi_{t-sq}\rangle=\sqrt{1-|\lambda|^2}\sum_{n=0}^\infty\lambda^n |n\rangle|n\rangle
\label{2-sq}
\end{equation}
which is the two-mode squeezed state with the squeezing parameter $\lambda$. When  $\lambda\rightarrow 1$, the two-mode squeezed state becomes a maximally entangled EPR state which was under scrutiny by Einstein, Podolsky and Rosen~\cite{EPR}. However such the state is unphysical as the mean energy of the state is infinite. Even though not maximal, it is clear that the two-mode squeezed state is entangled, but when it decoheres the state becomes mixed and entanglement is lost. In order to characterise quantum correlations of such the mixed state, criteria for quantum correlations were investigated~\cite{Reid,Simon}. In particular, the correlation studied by Reid is quantum steering, which has been generalised to other systems recently. Barnett and Knight~\cite{BK} studied the statistical properties of one mode of the two-mode squeezed state and found a thermal nature whose effective temperature grows with the degree of squeezing. This effect is closely connected to quantum entanglement of the two-mode squeezed state. Strong correlations in a two-mode squeezed state were achieved recently in~\cite{Schnabel} which reports as high as 10dB squeezing. Quantum correlation has been demonstrated beyond two-mode squeezing to establish entanglement in an ultra-large-scale network of quantum states, using SPDC and  beam splitters~\cite{Furusawa-multi}.  

While momentum and energy can have any values, the polarisation of a light field can have only one of the two values, horizontal/vertical or right/left circular. Understanding that the polarisations of the two SPDC photons are correlated, Aspect, Grangier and Roger showed the violation of Bell's inequality~\cite{Aspect}. Indeed, the polarisation degree of freedom having only two values shows that a photonic qubit can be realised by the polarisation of a single photon. Kwiat and his collaborators demonstrated how to generate an entangled polarisation state using Type II~\cite{Kwiat95} and Type I~\cite{Kwiat99}  SPDC processes.  
The polarisation entangled photon pairs have been used to demonstrate quantum teleportation~\cite{Bo}, entanglement swapping~\cite{Pan} and quantum key distributions~\cite{Erven}.  Since the bipartite entanglement was demonstrated, there have been an effort to increase the size of entangled photonic networks and it has been shown that as many as 8 photons are entangled~\cite{Yao12}. 

Even though an isolated two-level system such as an atom, an ion, a quantum dot or a defect in a diamond are all good candidates to generate single photons deterministically, SPDC has been at the forefront of the single photon generation, due to its simplicity. In many of the quantum information protocols, it is important to make sure that single photons are identical so that they interfere each other. A tool to make sure the single photon sources generate identical photons, the Hong-Ou-Mandel interferometer~\cite{Hong87} is used where the absence of coincidence counting rate will indicate how identical two photons are. Quantum interference has been generalised to a large number of input and output photons~\cite{Luca}. Recently technology has advanced to generate a few identical photons from quantum dots to interfere for their nonclassicality test~\cite{Somaschi,He}.

\subsection{Quantum state engineering}
The coherent states and squeezed states are represented by Gaussian Wigner functions in phase space. Beam splitting and phase shifting, which are common operations in an optical lab, are called Gaussian operations as they do not change the Gaussian nature of the input states. The homodyne and heterodyne measurements are then called as Gaussian measurements. Even though the Gaussian operations on Gaussian states are useful for certain tasks of quantum information processing such as continuous-variable quantum key distribution~\cite{QKD-Gauss} and precision measurements~\cite{Caves}, it has been found that Gaussian states may not be useful for other purposes of quantum information processing such as  simulations and computations~\cite{Eisert}. There have thus been studies on generating and manipulating non-Gaussian states~\cite{non-G}. 

Ourjoumtsev {\it et al} demonstrated a single-photon subtraction operation to convert a Gaussian state to a non-Gaussian one~\cite{Our}. Passing an initial field through a high-transmittivity beam splitter a photon is subtracted from the initial field, conditioned on the measurement/observation
 of a single photon in the other output of the beam splitter. The simplicity of the single-photon subtracting operation sparked a series of theoretical studies and experimental demonstration of using the operation to generate a superposition of two coherent states of a $\pi$ phase difference~\cite{Kim-cs,Grangier-cs} and to increase entanglement of weakly entangled two-mode squeezed state~\cite{Plenio-ent,Grangier-ent}. Vidiella-Barranco, Bu\v{z}ek and Knight~\cite{Buzek1} studied how the coherent superposition states decohere while the state has been shown useful for quantum computing~\cite{Kim-Jeong} and precision measurements~\cite{Munro}. 

Even though it is simple, a single photon subtracting operation does not bring a classical state into a non-classical one. On the other hand, single photon addition can convert any classical state into a non-classical one. Conceptually, single photon addition could be done by reverting the subtraction operation.  Instead of nothing impinged on the unused input port of the beam splitter, a single photon is impinged then a single photon is added on the initial state conditioned on measuring nothing at the other output mode of the beam splitter. This however requires mode matching between the single photon field and the initial field. Zavatta {\it et al.}  devised an ingenious scheme to use an SPDC process~\cite{Zavatta-a}  and converted a coherent state to a non-Gaussian state with a deep negativity in phase space. A sequence of single-photon additions and displacement operations can bring the vacuum state into an arbitrary quantum state written as
$
|\Psi\rangle=\sum_{n=0}^\infty c_n |n\rangle
$
where $c_n$ is complex~\cite{Welsch,Kim-review}. Another strikingly non-Gaussian class of states of the electromagnetic field are the Schroedinger cat states -- superpositions of quite distinct coherent states with a large negativity of the Wigner function. The coherence (quantified by the negativity of the Wigner function) and the decoherence (quantified by the disappearence of the same negativity) of such Schroedinger cat states of the electromagnetic field, and indeed their generation through dissipation, had been a topic of substantial interest in Knight's group~\cite{Garraway-Knight,Gilles}.

\subsection{Towards integrated quantum photonics}
One of the advantages of photons as a candidate to implement quantum technology is that their operations do not require cryogenic temperatures and they are hardly affected by environments. However, precise alignment of an optical setup maybe an issue regarding portability, integrability and scalability. Integrated photonic circuits (photonic chips) have been fabricated to put all the basic optical elements into a small chip to overcome these problems. As was shown by Reck {\it et al.}~\cite{Reck}, any discrete
qubit operation  can be done by a set of beam splitters and an arbitrary reflectivity beam splitter can be devised by a Mach-Zehnder interferometer (MZI). A MZI is a building block for arbitrary qubit operations. We can also combine beam splitters and photodetectors to realise two qubit gate operations such as the controlled-phase operations proposed by Knill, Laflamme and Milburn (KLM)~\cite{KLM}. There have been a good progress in fabricating photonic chips even to accommodate polarisation state~\cite{Rome} of light and photodetector units.  The photodetector is also an important component. At the moment, the most efficient detection of photons is achieved by using the change of resistance of superconducting wires~\cite{Nam}. The effort in building photonic chips has been extended to integrate photon sources as they implant the nonlinear materials on the chip for on-chip SPDC. Using the photonic chips, various tasks of photonic quantum information processing  have been demonstrated including boson sampling~\cite{Walmsley-BS}.   

As seen above, photons are useful for quantum technology for computation and simulation and information carrier for long-haul communications. The only way to realise long-haul QKD is using photons. Even though photons are relatively robust against environmental effects~\cite{QKD-Gauss}, they still decohere and lose entanglement and nonclassical effects. Even though it is not a pre-requiste, quantum repeaters~\cite{q-repeater} and quantum memories~\cite{q-memory} are useful. 
 
 \section{Cavity QED}
 \label{jcm}
The JCM has been one of the important tools to characterise the quantum-mechanical interaction of an atom with a single-mode field~\cite{Knight2} where the revival of Rabi oscillations was an important manifestation of the quantisation of the field, first discussed by Eberly et al~\cite{Eberly}. Knight and Radmore studied the revival when the field is initially in a thermal state~\cite{Radmore}.  An important experimental advancement was made by being able to prepare the atom initially in their superposition state~\cite{Haroche-d}. This opened up a possibility to exploit quantum coherences in cavity QED, which has enabled a way to reconstruct the cavity field~\cite{Haroche} and to produce a large superposition of coherent states in a cavity~\cite{Haroche-d,Buzek2}. In particular, a complementarity test in the Einstein-Bohr dialogue was experimentally demonstrated, based on entanglement between the atom and the cavity field~\cite{Haroche-c}. 

When a three-level atom interacts with a cavity field, a superposition state can be generated in the cavity whose nonclassical characteristics were investigated~\cite{W}. One of the first information theoretic studies of the atom-field interaction was done by Phoenix and Knight~\cite{Knight5} which introduces the concept of entropy to investigate quantum interaction. The time-dependence of the field entropy shows the evolution of entanglement between the two interacting sub-systems. Knight and colleagues~\cite{Bose99,Plenio99} proposed to entangle two atoms sitting in their respective cavities and to teleport the state of an atom from one cavity to the other using the cavity decay, which triggered a discussion on distributed quantum computation between two remote locations using photons as a messenger.

We can generalise the theory of cavity QED with a spin-boson interaction model as the cavity field is bosons and the atom is equivalent to a spin. The spin-boson model has been realised in various experimental setups such as superconductor circuit QED, ionic motion in a harmonic trap and a nanodiamond in a trap. Nonclassicalities of the ionic motion were demonstrated experimentally which has become a basis of an ionic quantum computer~\cite{Wineland1,Wineland2}. Knight and coworkers had taken a slightly different angle to propose quantum computation using vibrational coherent states~\cite{Knight-v}.

\section{Trapped ions}
\label{ion-traps}

Trapped ions are among the most fascinating systems, as experiments with individually trapped ions managed to contradict Schr\"odingers famous prediction that we would never perform experiments with individual quantum objects~\cite{Schroedinger:1935lr}.
Ion traps permit to confine charged atoms so well that the light emitted by one individual atom can be recorded~\cite{PhysRevA.22.1137} and that the properties of emitted light permit to draw conclusions on the electronic state of the ion~\cite{PhysRevLett.57.1699}.

The prospect of working with individual quantum objects instead of an ensemble and to probe their properties with unprecidented level of accuracy has resulted in concerted efforts to manipulate trapped ions for the exploration of fundamental physics and, more recently, for the development technological applications, in particular, quantum information processing.

The motion of ions in real space can be manipulated through electric fields.
Since the electric potential in vacuum satisfies $\Delta\Phi=0$ due to Gauss' law, there is no potential $\Phi$ that is attractive in all three spatial direction, but there is always at least one spatial direction in which the potential is repulsive.
Overall attractive potential can, however, be realised through a combination of static electric and magnetic fields or through temporally modulated electric fields.
Traps based on the former principle are called Penning traps and traps based on the second principle are called Paul trap or radio-frequency trap Paul trap~\cite{RevModPhys.62.531}.
Given the unique opportunities to manipulate and probe individual quantum particles, trapped ions have developed into an excellent testbed for fundamental physics~\cite{doi:10.1080/001075197182540} and quantum information processing~\cite{blatt_physrep}.

Before coherent operations on individual ions can be performed, the motion of the ions needs to be cooled close to its quantum mechanical ground state.
As established by equilibrium thermodynamics, energy flows from cold to hot systems,
what poses a fundamental challenge for the cooling of trapped ions to temperatures in the $\mu K$ regime in a room temperature environment.
Laser cooling~\cite{RevModPhys.58.699} 
overcomes these difficulties and permits to cool trapped ions close to their quantum mechanical ground state~\cite{PhysRevA.25.35,PhysRevLett.62.403}.
Since cooling is necessarily an incoherent process, it is typically not applied during coherent gate operations.
But, given the enormous temperature difference between the trapped ions and their surrounding, the ions motions tends to be heated up as soon as there is no active cooling present.
Such heating and resulting decoherence limits in particular the life-time of non-classical motional states~\cite{PhysRevA.58.663}.
Whereas ion traps of the early generations were macroscopic in size, there has been a trend of miniaturisation such that ions are trapped in proximity to trap electrodes.
The correspondingly increased interaction with noisy electric fields makes heating a highly relevant obstacle to coherent operations of trapped ions~\cite{PhysRevLett.97.103007}.
There is good understanding of the microscopic mechanisms responsible for heating ~\cite{PhysRevA.84.023412},
but the best methods to reduce heating is still simply enlarging the distance between ions and trap electrodes~\cite{PhysRevLett.116.143002}.
The heating rates of state of the art experiments are low enough for many proof of principle experiments, but heating will be a bottleneck for large scale quantum information processing requiring stable phase coherence over long times. Nevertheless, as we will explore further below, ion traps form one of the leading candidates for scalable quantum computing.

\section{Optomechanics: Quantum Optics with Mechanical Oscillators}\label{opto}
We have discussed in section \ref{jcm} how light can couple to multi-level systems such as atoms to provide interesting nonlinear dynamics in the fields of cavity-QED and circuit-QED. Through those internal variables they can also couple to the motion of the atoms as discussed in the section \ref{ion-traps} on trapped ions. However, light can couple to the motional variables of a material object more {\em directly} as well: through its radiation pressure on the object~\cite{Pace-Walls} or through the polarizability of the material~\cite{Chan}.  In both the above cases, the fundamental light-matter coupling is of an interesting ``tri-linear" form -- the photon number in the cavity (quadratic in creation/annihilation operators)
couples to a position (linear) variable of the mechanical object all these operator are linear.
Let us assume that the mechanical object is  a quantum harmonic oscillator with frequency $\omega_m$ 
and annihilation operator denoted by $b$, interacting with 
a cavity field mode of frequency $\omega_0$ and annihilation 
operator denoted by $a$. The relevant full opto-mechanical Hamiltonian~\cite{Bose-Jacobs-Knight-1997} 
is
\begin{equation}
\label{g}
\hat{H}_{\text{opto-mech}} = \hbar\omega_0~\hat{a}^\dagger \hat{a} ~+~\hbar\omega_m~\hat{b}^\dagger 
\hat{b}~-~\hbar g~\hat{a}^\dagger \hat{a} (\hat{b}+\hat{b}^\dagger)
\end{equation}
where the light-matter coupling of strength $g$ is called the single photon nonlinearity as it quantifies the effect of a single photon on mechanics. When the mechanical object is one of the mirrors of a Fabry-Perot cavity (assuming the mirror is movable and a harmonic oscillator) then it can be shown, by considering an adiabatic (slow) movement of the mirror that 
\begin{equation}
\label{gpar}
  g_{\text{mirror}}~=~\frac{\omega_0}{L} ~\sqrt\frac{\hbar}{2m\omega_m} ~,
\end{equation} 
and  $L$ and $m$ are the length of the cavity and mass of 
the movable mirror respectively. In the case that the movable object is a trapped mechanical object (we call this a ``bead") inside a cavity with static mirrors, the coupling is given by~\cite{Chan}
\begin{equation}
g_{\text{bead}} = \frac{3V_{\text{bead}}}{4V_{\text{cavity}}}\frac{\epsilon -1}{\epsilon +2} \omega_0,
\end{equation}
where $V_{\text{cavity}}$ is the cavity mode volume, $V_{\text{bead}}$ is the volume of the trapped bead and $\epsilon$ its electric permittivity.  

The above Hamiltonian with its tri-linear coupling is very interesting from the point of view of constructing non-classical states of both the electromagnetic field and the material object. The amount of non-linearity or the amount of departure from Gaussianity after starting from Gausian initial states in the evolution obviously depends on the ratio $g/\omega_m$, i.e, the relative strength of the trilinear term with respect to the harmonic term of the mechanical oscillator. Considering a regime when $g\geq \omega_m >> \kappa_c$, where $\kappa_c$ is the rate of decay of the cavity field,  Knight and co-workers~\cite{Bose-Jacobs-Knight-1997}, in parallel with~\cite{Mancini-Manko-Tombesi}, were one of the earliest teams to show how non-classical non-Gaussian states such as  Schr\"odinger
cat states of the light field could be produced as a result of the dynamics stemming from the above optomechanical Hamiltonian $\hat{H}_{\text{opto-mech}}$. Additionally, Knight and co-workers showed that the same Hamilonian enabled the generation of non-classical states of the mechanical object. Such states are known to be very important for quantum technologies, such as accelerometry~\cite{Kurt-2016}. However, except a couple of upcoming systems~\cite{Murch,Johansson,Levitated}, the above regime has not been reached till date. Thus most experimental results in optomechanics have thus, to date, been obtained in the presence of a strong driving of the electromagnetic field so that an effective coupling becomes strong at the expense of becoming linearized. We will discuss this regime next in short before proceeding to the details of nonlinear optomechanics.
\subsection{Optomechanics: Linearized Regime}

 When the electromagnetic field mode in a cavity is strongly driven such that there is
a non-zero amplitude $\alpha$ of the field in the cavity, one can rewrite the Hamiltonian of Eq. (\ref{g}) in terms of dispaced creation/annihilation operators $\hat{a} \rightarrow \hat{a}+\alpha, \hat{a}^{\dagger} \rightarrow \hat{a}^{\dagger}+\alpha^{*}$. Then the third (interaction) term of the Hamiltonian becomes the effective linearized optomechanical interaction 
\begin{equation}
\hat{H}_{\text{Linearized}}= g_{\text{eff}} (\hat{a}+\hat{a}^{\dagger})(\hat{b}+\hat{b}^{\dagger}),
\end{equation}
where $g_{\text{eff}}=g|\alpha|^2$ is the new (effective) optomechanical coupling strength which is much stronger the bare coupling strength $g$. From the above Hamiltonian, and assuming that the laser of frequency $\omega_L$ driving the cavity is detuned by $\Delta$ from the cavity frequency $\omega_0$, then three distinct quadratic Hamiltonians can be generated in accordance to the amount of detuning~\cite{Marquardt} in the interaction picture by eliminating terms rotating much faster than others 
\begin{eqnarray}
\hat{H}_{\text{Beam-Split}}&=&g_{\text{eff}} (\hat{a}\hat{b}^{\dagger}+\hat{a}^{\dagger}\hat{b}), ~(\text{Red detuned:}\Delta=-\omega_m)\nonumber\\ \hat{H}_{\text{Ent}}&=&   g_{\text{eff}} (\hat{a}^{\dagger}\hat{b}^{\dagger}+\hat{a}\hat{b}), ~ (\text{Blue detuned:} ~\Delta=\omega_m)\nonumber\\ \hat{H}_{\text{Non-demol}}&=& \hat{H}_{\text{Linearized}}. ~ (\text{Resonant case:} ~\Delta=0)
\end{eqnarray}
One can physically understand the red/blue detuned Hamiltonians from the fact that a phonon (mechanical oscillator quantum) has to be subtracted/added from/to the mechanical mode in order for the cavity mode to be populated by the driving laser.
The above provides a rich tool-box for Gaussian operations on mechanical and light modes. For example, the beam splitter interaction $H_{\text{Beam-Split}}$ can be used to swap the state of a mechanical object with the displaced light mode in the cavity. This can be used (i) for transferring the thermal quanta from the mechanics to light in the a cavity, which subsequently leaks out; this is thereby a mechanism for cooling the (potentially macroscopic) mechanical oscillator -- this is called cavity cooling, (ii) for creating an interesting non-classical, potentially non-Gaussian state of the mechanical object by injecting an optical mode in such a state into the cavity and then swapping its state with the mechanical object~\cite{Chan}. Additionally, as mechanics  can generically interact with most quantum systems, such as two-level systems or electromagnetic fields over wide range of frequencies, $\hat{H}_{\text{Beam-Split}}$ finds application in enabling transducers between quantum systems which would not naturally interact~\cite{Zoller-transducers}.  $\hat{H}_{\text{Ent}}$ is the well known two mode squeezing Hamiltonian, and is known to  create continuous-variable EPR entangled state of the two interacting modes -- in this case one of the modes is electromagnetic, while the other is mechanical~\cite{Tombesi-Vitali}. $\hat{H}_{\text{Non-demol}}$ is particularly useful to perform a non-demolition read out  the position $\propto (b+b^{\dagger})$ of the mechanical oscillator through a homodyne measurement on the electromagnetic field~\cite{Braginsky}. This can also be used to perform a full state reconstruction of the mechanical oscillator through direct position measurements by strong pulse of light~\cite{Vanner}.

\begin{figure}
\includegraphics[width=6cm]{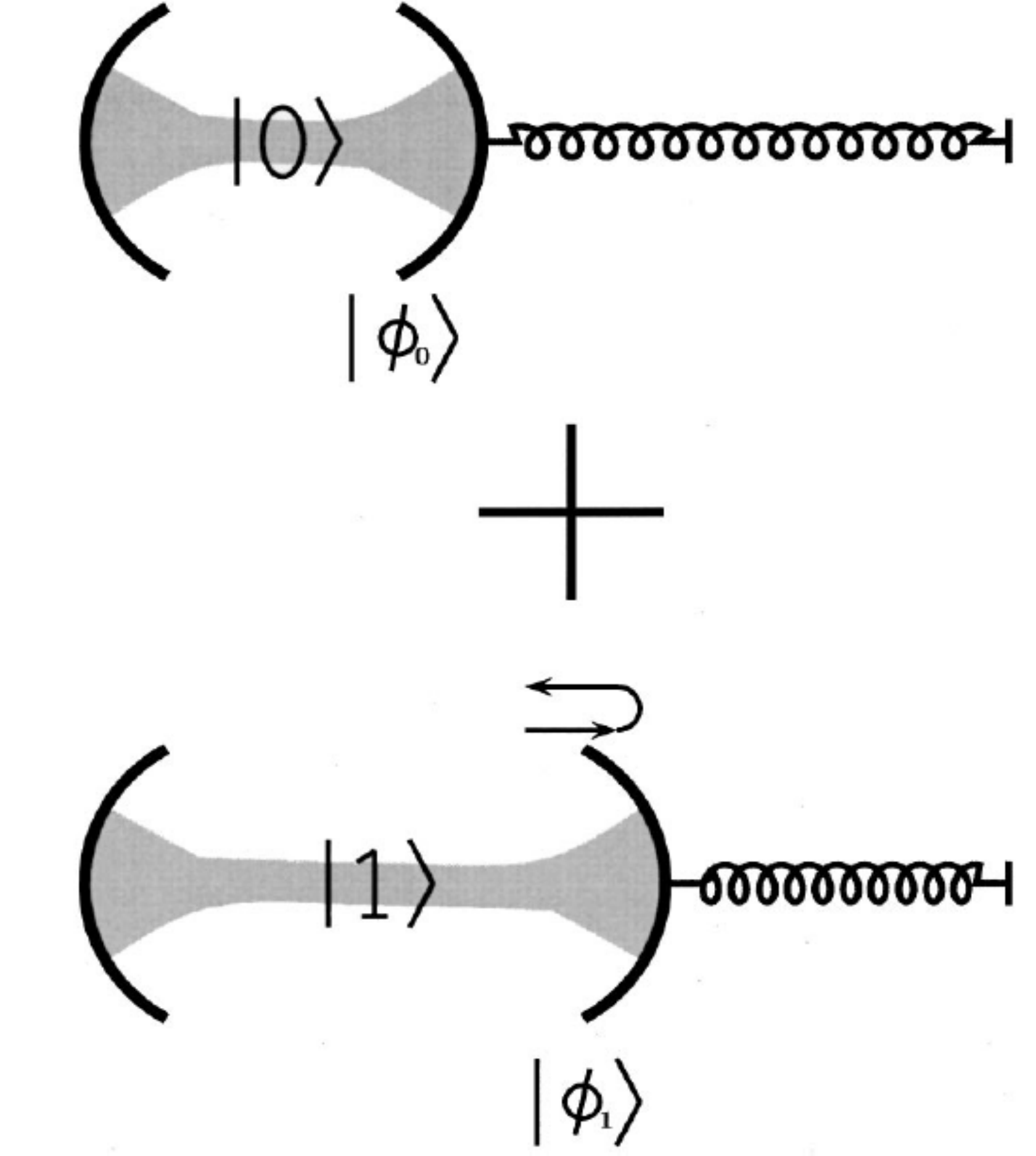}
\caption{\label{optomechfig} A superposition of Fock states $0\rangle$ and $|1\rangle$ is created inside a cavity. The mirror oscillates about different equilibrium separations corresponding to each Fock state, thereby creating a superposition whose components contain distinct coherent states at scaled times $t\neq 2m\pi$ (where $m=$integer). The coherence between the components of this superposition can be certified by probing the coherence between the cavity states $|0\rangle$ and $|1\rangle$ at time $t=2\pi$. \copyright American Physical Society~\cite{Bose-Jacobs-Knight-1999}.
}\end{figure}

\subsection{Nonlinear Regime:}  If one intends to go beyond the domain of Gaussian states of the mechanics and the electromagnetic field using optomechanics alone, one has to use the Hamiltonian of Eq.(\ref{g}) with the full tri-linear optomechanical interaction. 
 The explicit time evolution by assuming an initial state at time $t=0$ such as  
\begin{equation} 
  |\Psi(0)\rangle = |\alpha\rangle_{\mbox{\scriptsize c}} \otimes |\beta\rangle_{\mbox{\scriptsize m}}
\end{equation}
where $|\alpha\rangle_{\mbox{\scriptsize c}}$ and 
$|\beta\rangle_{\mbox{\scriptsize m}}$ are initial coherent 
states of the field and the mirror respectively, is given in the interaction picture by~\cite{Bose-Jacobs-Knight-1997}
\begin{eqnarray}
\label{ro1}
  |\Psi(t)\rangle~=~e^\frac{-|\alpha|^2}{2}\sum_{n=0}^\infty
\frac{\alpha^n}{\sqrt{n!}}e^{ik^2 n^2 (t-\sin{t})}|n
\rangle_{\mbox{\scriptsize c}} \otimes |\phi_n (t)\rangle_{\mbox{\scriptsize m}}
\nonumber\end{eqnarray}
where $|n\rangle_{\mbox{\scriptsize c}}$ denotes a Fock 
state of the cavity field with eigenvalue $n$, and the 
$|\phi_n (t)\rangle_{\mbox{\scriptsize m}}$ are coherent 
states of the mechanical oscillator given by
\begin{equation}
\label{phi1}
  |\phi_n (t)\rangle_{\mbox{\scriptsize m}} =|~\beta e^{-it} 
+ kn(1-e^{-i t})~\rangle_{\mbox{\scriptsize m}} ~, 
\end{equation}
in which $k=g/\omega_m$, and $t$ stands for scaled time $\omega_m t$.
We see that after a time $t=2\pi$ the mechanical oscillator returns to its original 
state. There is now an effective Kerr nonlinearity $\propto (a^{\dagger}a)^2$ on the cavity field 
so that physically one might expect the cavity field to have an 
evolution similar to that seen under a  Kerr like nonlinearity, namely the generation of Schroedinger cat states. The physical reason is that the position of the mirror  $\propto (\hat{b}+\hat{b}^{\dagger}) \propto \hat{a}^{\dagger} \hat{a}$, so that the effective Hamiltonian on the cavity has a term $\hat{a}^{\dagger} \hat{a} (\hat{b}+\hat{b}^{\dagger}) \propto (\hat{a}^{\dagger} \hat{a})^2$. Indeed, this is the case. For example, when the cavity mode and the mechanics disentangle at time $t=2\pi$, then for $k=1/(2\sqrt 2)$, we get a four component cat state
\begin{equation}
  |\zeta_4\rangle_{\mbox{\scriptsize c}} = \frac{e^{i\frac{\pi}{4}}}{2}(|\alpha\rangle_{\mbox{\scriptsize c}} - |-\alpha \rangle_{\mbox{\scriptsize c}}) +  \frac{1}{2}(|i\alpha \rangle_{\mbox{\scriptsize c}} +|-i\alpha \rangle_{\mbox{\scriptsize c}}).
\end{equation}
Thus by adjusting the ratio of the bare opto-mechanical coupling and the mirror frequency and thereby varying $k$ one can, in principle, obtain all these types of cats at time  $t=2\pi$. 

It is even more striking to note that at all times between $t=0$ and $t=2\pi$ the mechanical oscillator state  
is {\em entangled} with the field state with the entanglement 
being maximum when $t=\pi$. This entangled state is manifestly of non-Gaussian nature as it's superposed components
have distinct coherent states and there is a {\em periodic entanglement-disentanglement} dynamics. It was shown by Knight and co-workers that this dynamics is very suitable for probing the coherence between superpositions of distinct states of a macroscopic mechanical oscillator~\cite{Bose-Jacobs-Knight-1999} wherein only the electromagnetic field coupled to it needs to be measured. This scheme is depicted in Fig.\ref{optomechfig}. This basic methodology of probing the macroscopic limits of the quantum superposition principle has been adopted and incorporated into various subsequent schemes~\cite{Blencowe,Bouwmeester,Scala-2013,Wan-2016}.


\section{Quantum Technologies}\label{QT_sec}

\subsection{Quantum simulators} It is only in the last couple of decades that such simulators have been proposed~\cite{Jaksch-Zoller} and realized~\cite{Bloch} for specific idealized models of many-body physics (whose dynamics are still very difficult to predict under general conditions). Initially and primarily implemented with ultra-cold atom systems~\cite{Jaksch-Zoller,Duan-Demler-Lukin,Bloch,Greiner}, other avenues such as trapped ions~\cite{Porras-Cirac,Mintert-Wunderlich,Richerme}, superconducting arrays~\cite{Martinis-Chiral}, integrated photonic chips~\cite{Sciarrino-Fazio}, coupled matter-light systems~\cite{Angelakis,Hartmann-Plenio,Greentree,Cho,Houck}, and, most recently, solid state platforms~\cite{Rogge,Vandersypen,Gray,Banchi}, are also being studied. These simulators, where the interactions between the components are hard-wired to realize a specific physical model, are called ``analog" quantum simulators. There is, of course, also the effort to build a universal quantum computer in many platforms as described in section \ref{QC}. Such machines can be programmed to mimic a wide variety of distinct (ideally any) complex quantum systems and phenomena therein. Here multiple pulses may be used to control the systems according to the program and realize a complex evolution in terms of a series of several shorter evolutions~\cite{Trotter-paper}. Each of these smaller unitary evolution steps can be thought of as a quantum gate, with the whole simulation being a quantum circuit. Because of the quantum circuit methodology for implementation, in principle, quantum error correction can be incorporated to make such simulators highly accurate.  Such simulators are called ``digital" quantum simulators, and although current implementations are still without error correction, these have been already impemented in superconducting qubit architectures ~\cite{Wallraff-Solano,Martinis-Solano} and photonic chips~\cite{Sciarrino-Banchi}.

We start from the area of ultra-cold atoms where essentially a Bose-Einstein condensed gas of atoms is subjected to a lattice potential created by light. The light is highly detuned from the internal atomic resonances and creates an intesity dependent periodic potential which is largely conservative. Because of their very small masses, atoms (an atom used typically is Rubidium) can tunnel over micron separations between adjacent potential wells of the lattice. Moreover, two atoms in a given lattice site can interact with the contact interaction, which can be enhanced via a Feshbach resonance. The tunneling energy $t$, the atom-atom on-site interaction energy $U$ and the chemical potential (the energy of a single atom placed in a site) $\mu$, each in frequency units, together give rise to the Bose-Hubbard Hamiltonian~\cite{Jaksch-Zoller}
\begin{equation}
\hat{H}_{\text{BH}}=t\sum_{\langle ij\rangle} (\hat{a}_i^{\dagger} \hat{a}_j  + h.c) + \sum_i U (\hat{a}_i^{\dagger} \hat{a}_i)(a_i^{\dagger} \hat{a}_i-1)+\mu \sum_i \hat{a}_i^{\dagger} \hat{a}_i,
\end{equation}
where $\langle i,j \rangle$ represent nearest neighbour pairs of sites in the lattice and $\hat{a}_i^{\dagger}$ creates a particle in the $i$th site. Two stable phases of this model are the Mott insulator phase for $U>>t$ where there are a fixed and equal number of atoms in each site (depending on $\mu$) and the superfluid phase for $t>>U$ in which the superfluid order parameter $\langle a_i\rangle$ assumes a non-zero value. In a remarkable experiment it was shown that one can observe a quantum phase transition from one of these phases to the other as the ratio $t/U$ is varied~\cite{Bloch}. Similarly experiments have now been performed with fermionic atoms~\cite{Esslinger}, where the phase transition is from a metallic to a Mott phase. This opens the door to controllably simulate, and thereby understand, emergent many-body phenomena in condensed matter such as high temperature superconductivity which are thought to take place in doped Mott insulators. Moreover the Mott insulator phase with exactly one atom per site gives rise an antiferromagnetic Heisenberg spin model for pseudo-spins represented by two internal levels of an atom. This is because the atoms, being prohibited to doubly occupy a site, can only swap their positions through a second order process with frequency $J\propto t^2/U$, which is equivalent to swapping their pseudo-spins with their positions remaining fixed, or, in other words, the isotropic Heisenberg Hamiltonian
$\hat{H}_{\text{Heis}}=J(\hat{X}\otimes \hat{X}+ \hat{Y}\otimes \hat{Y}+\hat{Z}\otimes \hat{Z})$ (here $\hat{X},\hat{Y}$ and $\hat{Z}$ are the Pauli matrices) acting between their pseudo-spins. With internal level dependent hoppings, various anisotropic Heisenberg models can hence be synthesized~\cite{Duan-Demler-Lukin}, thereby also opening the door for exotic spin liquid phases to be realized. The same philosophy with $N$ degenerate
 atomic levels as in alkaline earth metals, leads to models with $SU(N)$ symmetry~\cite{Hadley,Gorshkov}, and, eventually, simulators for strongly coupled quantum field theories such as quantum chromodynamics which are difficult to solve~\cite{Zoller-QCD}. 

Note that $\hat{H}_{BH}$ is essentially obtained by adding an on-site nonlinearity (a Kerr nonlinearity in quantum optics terms) to a free (quadratic) model (hopping model). It is quite natural that the idea of an analogous model for photons, where they hop between distinct electromagnetic cavities, and a  nonlinearity provided by two-level systems (e.g., atoms) trapped in each cavity, will originate from  former associates and students of Peter Knight~\cite{Angelakis,Hartmann-Plenio}. In parallel with another independent group~\cite{Greentree}, they have introduced the so called Jaynes-Cummings-Hubbard model, whose ideal realization seems to be with connected microwave cavities, each interacting with supeconducting qubits~\cite{Houck}. This uses the basic fact that the JCM (section \ref{jcm}) has an anharmonic level structure so that multiple polaritonic excitations (atom plus photon combined excitations) at a site costs more energy than individual excitations. Accordingly, these models were also shown to exhibit a Mott-superfluid-like quantum phase transition~\cite{Angelakis,Hartmann-Plenio,Greentree}, spin models with atomic internal levels acting as pseudo-spins~\cite{Angelakis}, and, from thereon, even fractional quantum Hall states using two dimensional arrays of connected cavities~\cite{Cho}.

Ion traps have provided a fertile ground for quantum simulation experiments. The harmonic degrees of freedom are ideally suited to mimic the behaviour of light,
and ideas to simulate optical elements~\cite{PhysRevA.54.1682} and cavity electrodynamics ~\cite{PhysRevA.56.2352} were among the first proposals for quantum simulations with trapped ions. Among the numerous ideas for quantum simulation of more complex systems,
are relativistic dynamics~\cite{PhysRevA.76.041801} and phase transitions in anharmonically interacting many-body systems~\cite{PhysRevLett.101.260504}.

Proposals focussing on the qubit-like degrees of freedom, suggested trapped ions as quantum simulator of  spin models~\cite{Mintert-Wunderlich,0953-4075-36-5-325,Porras-Cirac} of molecules or solid-state systems.
Subsequent developments included the realisation of more than pairwise interactions~\cite{PhysRevA.79.060303},
or localisation phenomena induced by disorder~\cite{1367-2630-12-12-123016}.
With ideas to simulate frustration in interacting quantum-spins~\cite{PhysRevLett.107.207209} and magnetism as induced by artificial magnetic fields~\cite{PhysRevLett.107.150501},
the theoretical footing for quantum simulations are sufficiently advanced to use trapped ions for simulations that exceed the capacities of available classical computers.
As a result, trapped ions are nowadays a popular platform for proposals to use quantum simulations for studies that suffer from limitations in our computational resources.
Among the more recently developed ideas are the application of two-dimensional spectroscopy to trapped ions~\cite{1367-2630-16-9-092001,PhysRevA.90.023603,PhysRevLett.114.073001}.

 Although the spin-spin coupling simulated in an ion trap can be uniaxial/Ising, e.g., $\hat{Z}\otimes \hat{Z}$~\cite{Mintert-Wunderlich,0953-4075-36-5-325,Porras-Cirac}, either a strong effective field in another direction (which could be a laser driving)~\cite{Richerme} or a stroboscopic (digital simulator) method whereby the spin bases are changed regularly from $\hat{Z}$ to $\hat{X}$ and/or $\hat{Y}$ by appropriate pulses~\cite{Wunderlich-Illuminati}, can be used to generate more general Hamiltonians. 
Similarly, superconducting qubit arrays naturally have either Ising couplings through direct (say, capacitive) interactions between neighbouring qubits or $\hat{X}\otimes \hat{X}+\hat{Y}\otimes \hat{Y}$ couplings when the interactions are mediated by an adiabatically eliminated microwave bus. Nontheless, these have been shown to be versatile for digital simulations of generic spin models using the technique of basis rotation by regular pulses~\cite{Wallraff-Solano} and even for the digital simulation of itinerant fermionic particles by exploiting the Jordan-Wigner transformation mapping spins to fermions~\cite{Martinis-Solano}.  Different simulators have different advantages and limitations. For example, measurements of long range correlations and entanglement may be difficult for the atomic and solid state simulators as interactions typically fall with distance, but photonic chips can bring very distant photons to interfere at a common beam splitter -- this has been used to verify the dynamics of long range entanglement growth following a quench in an emulated $\hat{X}\otimes \hat{X}+ \hat{Y}\otimes \hat{Y}$ spin model~\cite{Sciarrino-Banchi}.  Here it is worth mentioning that the quantum simulators are also opening up new vistas: both ultra-cold atomic and ion-trap simulators are very well isolated from their environments unlike typical solid-state many-body systems of nature. They thus enable the study of long time nonequilibrium dynamics (more precisely unitary evolution) of a non-eigenstate of a many-body Hamiltonian. This in turn enables exploring fundamental phenomena such as dynamics of excitations~\cite{Richerme,Fukuhara,Lanyon} (which can, in turn, have quantum technology applications~\cite{Bose-2007,Compagno}), as well as probe notions of statistical physics  such as ergodicity and localization in the quantum regime~\cite{Bloch-MBL,Monroe-MBL}.

It is worthwhile stressing that one of the fundamental lessons that quantum simulators teach us is that physical systems which are apparently quite disparate can be made to mimic each other. On this aspect we point out Peter Knight's work on the cavity-QED and optical implementations of coined quantum walk~\cite{Knight-Sanders-walk,Knight-Roldan}, which can both simulate the same phenomenon  (a natural phenomenon which this mimics, on the other hand, is the spin dependent motion of a physical particle -- a Dirac-like Equation in the continuum limit~\cite{Meyer}).

\subsection{Quantum computing}
\label{QC}
Quantum computing is one of the most potentially transformative applications of quantum mechanics to technology. Representing data quantum mechanically allows one to compute in a fundamentally new way. Information can be stored in quantum degrees of freedom, such as the internal state of an ion, or the polarisation state of a photon, where coherence and entanglement allow for information processing in a fundamentally non-classical way.

The quantum bit or qubit is the basic building block of a quantum computer representing logical 0 and 1 in a pair of orthogonal quantum states $\ket{0}$ and $\ket{1}$.
In a quantum computer, a register of quantum bits stores information  in a coherent superposition of these states.
This information is manipulated via quantum logic gates, in analogy to the logic gates in a conventional ``classical'' computer. The most important logic gates include single qubit operations such as the Hadamard gate, which generates superposition states $\ket{0}\pm\ket{1}$ from the basis states $\ket{0}$ and $\ket{1}$ respectively. The controlled-NOT or CNOT gate is a two-qubit gate in which the state of the second qubit is flipped between $\ket{0}$ and $\ket{1}$, conditional on the state of the first qubit. In particular, CNOT gates can generate entanglement between the qubits.

Quantum algorithms for problems such as factoring numbers, have been shown to run with a run time which scales polynomially with the size of the problem (i.e. the number to be factored), while the best known classical algorithms have exponential run-time scaling\cite{Shor}. Quantum computers thus offer the potential to solve problems beyond the scope of conventional computing technology.

Any quantum system with two or more distinguishable states can represent a qubit and there are therefore a wealth of possible physical realisations of a quantum computer. The implementation of quantum computing remains an area of vibrant research and it is still not clear which type or types of physical system will be the most suitable platforms for developing  architectures for large-scale quantum computations.

Quantum optics and atomic physics research have delivered a number of candidate systems for quantum computing, including two of the currently  leading candidates, ion trap quantum computing, and linear optical quantum computing, which we will describe here. They have also inspired a third leading implementation, quantum computing with superconducting qubits, where superconducting circuits designed to mimic light and matter and its interaction, realise a rapidly developing technology for quantum computing in the solid state.

\subsubsection{Ion Trap Quantum Computing}
Ion traps represent one of the leading candidates for scalable quantum computing. A central aspect in the utilisation of trapped ions for quantum information processing, and also quantum simulations~\cite{RevModPhys.86.153} or sensing~\cite{PhysRevLett.116.240801}, is the ability to realise coherent gate operations of qubits, as typically encoded in electronic states of ions.
Realising single qubit gates is comparatively simple, but achieving entangling gates that require a coherent interaction is much more challenging.
Because of their mutual Coulomb repulsion, the distance between two trapped ions is too large for direct interactions (such as dipole-dipole interactions) to be sizeable,
and the electronic degrees of different trapped ions are essentially non-interacting.
It thus necessary to find an additional degree of freedom that can mediate an interaction,
and a mode of collective oscillation of a set of trapped ions, which is often referred to as bus mode is the most commonly employed solution.
Interactions between ionic qubits and a motional degree of freedom can be realised in terms of driving with a coherent light field of frequency close to resonance with the transition frequency of the two qubit states $|0\rangle$ and $|1\rangle$.
Since this transition is accompanied with the absorption or emission of a photon, and this photon has a finite momentum, the electronic transition also affects the motional state of a collection of ions.
Tuning the driving field exactly on resonance with the qubit transition permits to minimise this affect and to drive carrier transitions that leave the motional state of the ions invariant.
Choosing a detuning that matches the resonance frequency of the bus mode, however, makes the carirer transition energetically forbidden and results in a red or blue sideband transitions,{\it i.e.} a transition between the qubit states $|0\rangle$ and $|1\rangle$ and additional annihilation or creation of a phonon.

The first theoretical proposal~\cite{PhysRevLett.74.4091} to realise a CNOT gate is a concatenation of side-band and carrier transition on two different ions.
It requires the bus mode to be originally cooled to its ground state, so that the state in one qubit can be written in the bus mode with a red sideband transition.
A sideband transition (involving a third level) on a second ion induces a phase shift that is conditioned on the state of the bus mode, and thus on the original state of the first qubit.
Writing the state of the bus mode back to the qubit encoded in the first ion completes a controlled phase gate, which with some additional single qubit gates, realised through carrier transitions results in the desired CNOT gate.

The prospect of quantum logic inspired huge activities towards the realisation of controlled interactions,
both between motional~\cite{PhysRevA.56.4815}  and qubit degrees of freedom.
Whereas first ideas for the realisation of two-qubit gates were relying on a motional mode cooled to its quantum mechanical ground state and required substantial time for the exectution of a gate, the technical challenge of cooling trapped ions close to the ground state and to maintain phase coherence over long times was a strong motivation to develop ideas for fast gates with weak requirements to the properties of the bus mode.

A central step towards quantum gates that require no ground state cooling was the idea to realise the mediation of entangling interactions through Raman transitions in which changes in the motional state are only virtual~\cite{PhysRevLett.82.1835,PhysRevLett.82.1971}. Although the effective Rabi-frequency for sideband transitions depends on the phonon number, an interference of several path amplitudes makes the final gate independent of the initial motional state.
The entangling gate realised with this driving scheme thus even works for incoherent thermal initial states.
A practical limitation to low temperatures, however, is imposed by the fact that contributions of side-band transitions involving changes by more than one motional quantum need to be neglected which is valid only at sufficiently low temperatures.
The ability to realise entangling gates without the need to cool the ion's motion to its absolute ground state facilitated experiments substantially, and is certainly a central step from fundamental proof of principle experiments to a practical technology.
These advance also inspired substantial further theoretical work on refinement of gate operations,
in particular faster gates~\cite{PhysRevA.62.042307,PhysRevLett.87.127901}.

With many practical ideas for the realisation of quantum gates that are robust to systems imperfections,
it is certainly well established that trapped ions are suited for proof-of-principle demonstrations of quantum information processing.
The next logical step is the question of scalability, {\it i.e.} the realisation of hardware that can be assembled to a large computational unit with the potential to exceed the capacities of available classical computers.
Given the availability of flexible, high-quality microwave sources,
control of trapped ions without complex laser systems~\cite{PhysRevLett.87.257904,PhysRevLett.101.090502} is a route actively pursued towards the reduction of experimental complexity. Noise in the magnetic fields that are required to engineer interactions in the absence of laser-control, necessarily induce dephasing, but encoding of qubits in dressed states that are insensitive to such noise in leading order provides excellence phase coherence~\cite{nature:185.476.2011}. 

With the ability to implement quantum gates with sufficiently high fidelity, quantum error correction~\cite{PhysRevA.54.1098,Steane2551}
can be employed to improve the accuracy of quantum information processing.
Since quantum error correction requires encoding of a qubit in at least three ions, the experimental realisation of quantum error correction is rather challenging.
Once realised experimentally~\cite{Chiaverini:2004}, however, it got recognised as a very practical element in the transformation of an ion trap experiment to a quantum computer. With today's experimental abilities to control on the order of $10$ ions, also recently developed topological protection~\cite{PhysRevLett.97.180501} has been implemented~\cite{Nigg302}.

Theoretical contributions (beyond the development of the general framework of quantum error correction~\cite{PhysRevA.54.1098,Steane2551}) are mostly found in adapting general quantum error correcting schemes to the requirements of trapped ion experiments.
The modular structure of modern ion trap geometries suits the encoding of a qubit in several ions rather well~\cite{2017arXiv170205657L},
and the necessity to perform complex measurements is a bottleneck that can be overcome by simplified syndrom measurements in special geometric arrangement of trapped ions ~\cite{PhysRevA.92.032314}, or more autonomous solutions in which dissipation overcomes the necessity of active measurements and corrections~\cite{2017arXiv170208673R}.

Trapped ions are certainly a prime example of a quantum optical system that is on the threshold of being transformed into a technology.
The driving force behind the developments over the last decades was certainly the prospect of quantum computation, but also quantum simulations have established themselves as a large, active field.
While there has been tremendous efforts of reducing noise and decoherence in experiments, theoretical work has notably contributed to advances with ideas to realise desired coherent dynamics, even in experiments that do not achieve perfect suppression of imperfections. \\

 The prospect of quantum information processing certainly sparked enormous activities towards the development of ion trap technology that is capable of executing proper quantum algorithms. Given the enormous experimental challenge in the realisation of the desired technologies, theoretical developments are clearly exceeding the experimental progress. That is, despite the numerous theoretical ideas to realise and improve quantum gates, the experimental reality is still far away from a quantum computer exceeding the capacities of existing classical computers. 

Beyond exploring the potential of an ion trap quantum computer or quantum simulator,
there are also large theoretical efforts to support the experimental development of the necessary technology.
Since any hardware of practical technological value, requires a certain level of robustness against decoherence and other system imperfections, optimal control theory is a natural playing field in which theory can contribute to the development of experimental hardware.

Dynamical decoupling is successfully being applied for the effective reduction of coupling to environmental noise, and it can be utilised also to enhance the robustness of trapped ion quantum gates~\cite{PhysRevA.85.040302}.
For the degrees of freedom that encode qubits it seems natural to target decoupling from all sources of noise,
but for auxiliary degrees of freedom used to mediate interactions, this is not necessarily the only viable option.
Indeed, coherent quantum gates can also be mediated by dissipative motional modes,
either by minimising the impact of motional decoherence on entangling gates~\cite{1367-2630-18-12-123007} or by actively exploiting motional dissipation~\cite{PhysRevLett.110.110502}.

\subsubsection{Linear optical quantum computing}
Photons are natural candidates to represent quantum bits. For example,  states $\ket{0}$ and $\ket{1}$ may be represented as orthogonal polarisations (e.g. horizontal vs vertical) or by path degrees of freedom (which arm the photon travels down in an interferometer). 

Linear optical networks consisting of beam splitters and phase shifters, or  equivalently polarising beam splitters and polarisation rotators, allow all single qubit quantum logic gates to be realised. As has been discussed above, such networks also allow for arbitrary unitary mode transformations. It is therefore natural to ask whether universal quantum computing can be achieved by linear optics~\cite{kokrmp}.

To achieve a universal set of quantum gates requires the addition of a two-qubit gate such as the CNOT gate. Unfortunately, CNOT gates cannot be achieved via linear optical networks alone. Photonic modes interfere in linear optical networks, but photons do not directly interact. While linear optical transformations can lead to entanglement between modes (for example in Hong-Ou-Mandel interference, discussed above) it is impossible for a linear optical network to realise an entangling quantum gate on photon qubits.

Non-linear materials are a way to realise an effective photon-photon interaction and the Kerr effect can be the basis of  entangling photonic quantum gates~\cite{yamamotokerr,milburnkerr}.  Kerr non-linearities in natural materials are too weak for useful quantum gates to be realised. Methods that exploit atom-light interactions for stronger optical non-linearities remain an active area of research\cite{rempe-pi}. 

\begin{figure}
\includegraphics[width=6cm]{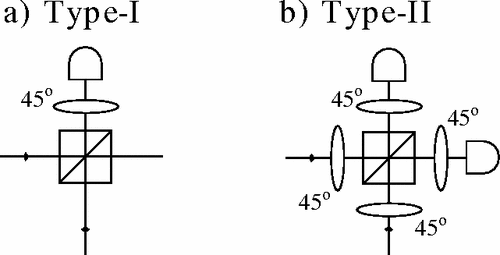}
\caption{\label{fig:loqc}
``Fusion gates''~\cite{browne_rudolph} realise a parity measurement~\cite{braunstein} via a single polarising beam splitter followed by measurement. They can be used to combine entangled multi-photon states into larger cluster states, resources for universal quantum computing via single-qubit measurement alone. \copyright American Physical Society~\cite{browne_rudolph}.
}\end{figure}

By combining linear optical networks with photon measurement, Knill Laflamme Milburn (KLM) cite{KLM} showed that entangling gates such as the CNOT gate could be realised, via a mechanism known
 as measurement-induced non-linearity~\cite{peter-meas-induced-non}. The entangling gates realised in this manner are non-deterministic, they only succeed some of the time, when a particular set of photon-detectors fire. Otherwise, the gate fails, destroying the quantum state (by measurement) of the qubits in the process.

This non-deterministic property means that one needs to take extra steps to achieve useful near-deterministic computation. KLM~\cite{KLM} proposed a method for increasing the success probability of the gate sufficiently high that error correcting codes could tolerate the remaining failure rate. This mechanism however, was very complicated and required a large number of additional photons to implement each CNOT gate.

Many alternative approaches have been proposed since this work, which significantly reduce this overhead. A popular approach is to use an alternative model of quantum computation, measurement-based quantum computation~\cite{mbqc}. Here, rather than realising quantum computations via quantum gates, equivalent computations are achieved via single-qubit measurements upon an entangled resource state known as a cluster state. Since linear optics and photon measurement allow measurement of single qubits in all bases, all that is then required for universal quantum computation is to create an entangled multi-photon cluster state. Nielsen~\cite{nielsenmbqc} showed that non-deterministic CNOT gates are particularly well suited to cluster states construction, reducing the overhead associated with KLM's approach significantly. 

Browne and Rudolph ~\cite{browne_rudolph} showed that Hong-Ou-Mandel type interference combined with photon-measurement provided a direct approach to constructing cluster states via so-called fusion gates, a form of parity meaurement~\cite{braunstein}, see figure~\ref{fig:loqc}. This reduced the resource overhead further, provided entangled photon Bell pairs and active switching of photons were available.

Many challenges remain for the large-scale implementation of linear optical quantum computation. It will require high efficiency sources of high quality single photons or (better) entangled photon pairs or triples, complex interferometers with some active switching and high efficiency detectors. Nevertheless, the key principles of all of these ingredients have been demonstrated, and linear optical quantum computing remains a strong candidate for the realisation of large-scale quantum computing~\cite{terryoptimistic}.

In the near-term, linear optics has become a strong candidate for the demonstration of so-called quantum supremacy, computational demonstrations of tasks which would be too hard for current classical computers to replicate. The demonstration of a  quantum algorithm of sufficient complexity might be one way to achieve this, but a potentially simpler alternative is offered by sampling problems. A sampling problem is a challenge to produce output bit-strings sampled according to a certain probability distribution. Properties of the distribution can make certain distributions hard to realise on a classical computer, and this hardness can be proven provided certain assumptions, which are widely believed by computer scientists, hold to be true.

Boson-sampling~\cite{aaronsonbosonsampling} is the name of a family of sampling problems defined by Aaronson and Arkhipov using linear optical circuits. The output of a circuit, with 
many single photons as input, and many single photon detectors at output can produce a distribution for which there is strong evidence that it is classically hard.  These arguments use elegant techniques from computer science, in particular computational complexity, but they can be understood to follow from a connection between linear optics and matrix permanents. A matrix permanent is a functional like the better known determinant, and is defined in a similar way, but without the alternating signs that characterise the determinant.  Unlike the determinant, for which there is an efficient algorithm, there is no known efficient classical algorithm for permanent computation and strong evidence that no such algorithm will be found. Scheel showed that the permanent arises naturally in the calculation of multi-photon states following a linear optical transformation~\cite{scheelpermanent}. The appearance of the permanent in linear optical network transformations is a signature of the potential for quantum supremacy later revealed by the Boson Sampling problem.

\subsubsection{Networked Quantum Computing}

While ion trap quantum computing and linear optical quantum computing each have distinct advantages, a further promising avenue is hybrid approach that combines features from each. Ion traps have the advantage of high-fidelity deterministic single qubit and multi-qubit gates. The principle challenge is in their scalability. Linear optical quantum computing, however, provides simple interferometric tools to generate qubit-qubit entanglement.

The key idea of networked quantum computation is to adopt a modular architecture~\cite{nqit1,nqit2}, where separate ion trap registers of a small number of qubits are linked together via optical fibres enabling measurement-induced non-linearities to create long-distance entanglement between separate ion trap registers. There are a number of potential advantages to this approach. The modular design should aid scalability, since the many ion trap registers could share a uniform design. 

\begin{figure}
\includegraphics[width=6cm]{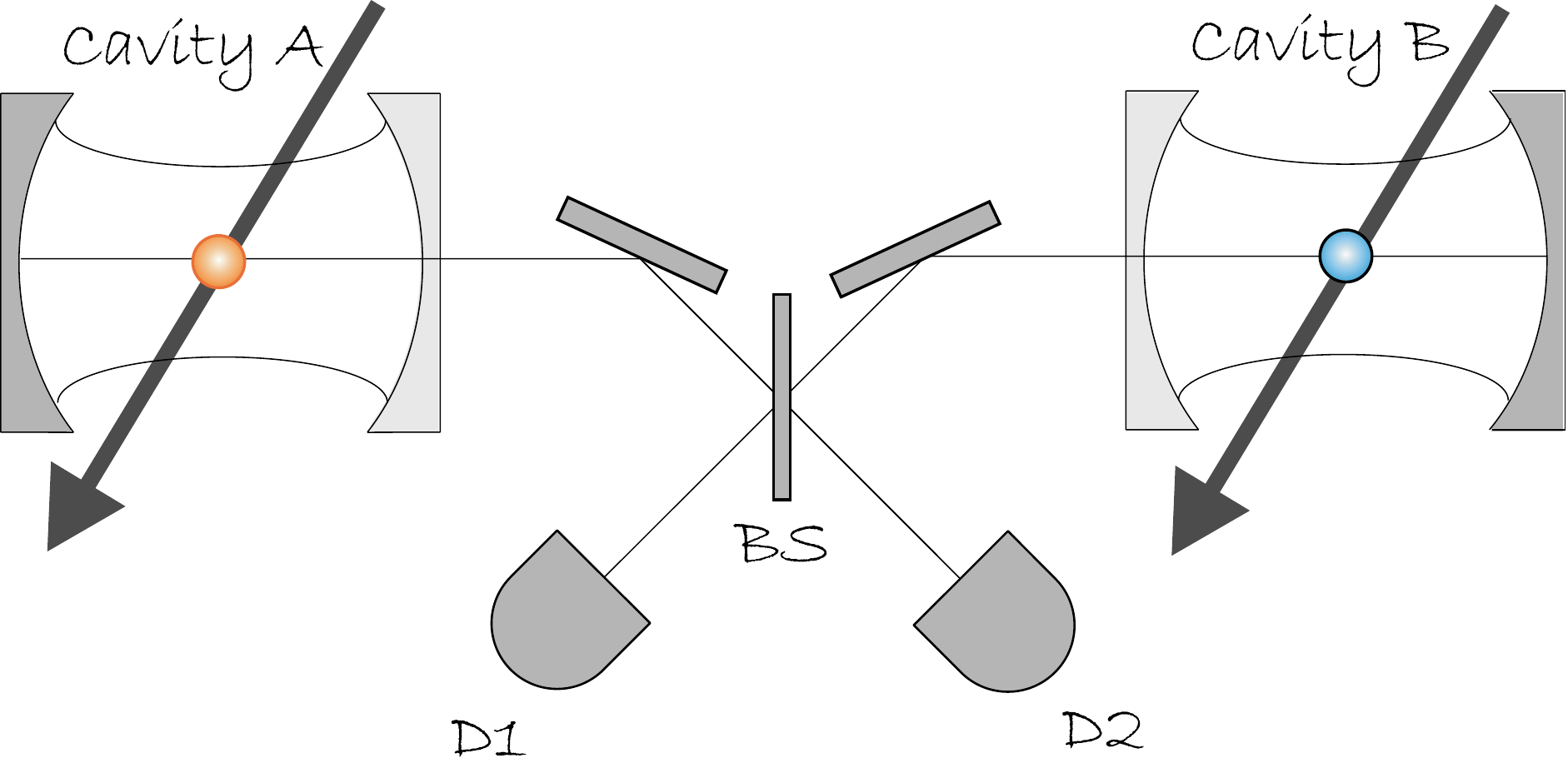}
\caption{\label{fig:cavity}
Spatially separated qubits (here ions trapped in separate cavities) can be entangled via the  emission of photons and interference on a beam splitter followed by photon detection. The erasure of which-path information on the beam splitter leads to the entanglement  \copyright American Physical Society~\cite{linoptentanglement_paper1}.
}\end{figure}

The idea of exploiting photon interference to entangle spatially separated qubits can be traced back to early work by the groups of Peter Zoller~\cite{cirac_lin_opts_entanglement} and Peter Knight ~\cite{knight_lin_opts_entanglement}, who showed that if objects emit
photons and those photons interfere, the erasure of which-path information due to the photonic interference can, after a subsequent  photon detection, allow for the generation of entanglement between the emitting objects. This is illustrated in figure~\ref{fig:cavity} Entanglement arises due to the erasure of which-path information together with the preservation of coherence in the process. 

Variations of this idea have been proposed which make the process more robust~\cite{linoptentanglement_paper1,linoptentanglement_paper2}, in particular, the proposal of Barrett and Kok~\cite{barrettkok}, which notably formed the basis of entanglement generation in the first loophole-free Bell inequality experiment~\cite{loophole_free_Bell}.

Realising such a networked approach to quantum computing is now one of the principle goals of the NQIT Networked Quantum Information Technologies Hub of the UK National Quantum Technologies programme, an ambitious and significant investment from the UK government towards the realisation of quantum technologies. The NQIT hub is developing a NQIT 20:20 prototype, where 20 ion trap modules will be connected via a linear optical network.
A device of this scale would represent a powerful demonstration of the scalability and potential of the networked approach to quantum computation.

\subsubsection{Superconducting qubits}

Superconducting qubits are one of the two leading approaches (together with spin qubits~\cite{kane}) for realising quantum computing in the solid state. We draw attention to superconducting qubits here as their design has incorporated a number of ideas from quantum optics.

Circuit QED is a direct analogue of cavity QED. In circuit QED~\cite{Wall}, a spin or an artificial atom is realised by the number of Cooper pairs in a superconducting island of the so-called Cooper-pair box which interacts with a microwave field in a superconducting coplanar waveguide resonator.  Recently, researchers have developed a circuit QED system less vulnerable to charge noise to increase the coherence time, which is called transmons~\cite{Koch}. The transmons are one of the strong contenders to realise qubits for quantum computing with a long coherence time and scalability. 

We can control the interaction strength of the circuit QED system. Indeed, the interaction strength can be made very large in comparison to the spin transition frequency. With the easy controllability of the interaction, the circuit QED system has realised some quantum simulation protocols. When the interaction strength is as large as the transition frequency, the rotating-wave approximation is not valid and the dynamics of the circuit QED system is described by the Rabi model rather than the JCM. Once the counter rotating terms are no longer neglected~\cite{Solano}, an infinitely large number of states become coupled and a quasi-equilibrium state appears even though the whole dynamics is described by a closed system~\cite{Hwang}.   

Transmon qubits have enabled the demonstration of a number of key elements of quantum computing, from long coherence times and the implementation of a universal set of gates~\cite{transmon1,transmon2,transmon3,transmon4},  to the demonstration of the building blocks of quantum error correction\cite{Chow_error,Martinis_error}.

\section{Conclusions and Outlook}
In the 1970s, when neither single atoms nor single photons, had yet been directly observed, a small group of physicists pioneered quantum optics. For the past 40 years, we have seen an enormous progress in understanding quantum mechanics and manipulating single atoms and single photons. Some popular textbooks still describe quantum mechanics as a law of nature at the subatomic scale. However, our ability to control nonclassical interactions between atoms and photons is getting finer and we are now capable of observing quantum behaviour in an object containing billions of atoms. Researchers are adopting quantum mechanics to tackle unanswered questions in various branches of science such as high-efficiency energy transport in photosynthesis. We are in an exciting era where our deepest  theories of physics can be tested using quantum optical tabletop experiment and where quantum mechanical ideas have found a central role in the development   of new technologies. 

The importance and potential of these developments has been recognised by the UK government with a significant investment, through EPSRC, in the National Quantum Technologies programme. Peter Knight played a crucial role in convincing the UK government to support this ambitious programme and helped shape its activities on its Strategic Advisory Board and beyond.

Research in this programme is led by  four {\em Quantum Technology Hubs}. A hub centred at Birmingham University is developing quantum enhanced {\em Sensors and Metrology}, while the {\em QUANTIC} hub lead by Glasgow University is working towards {\em Quantum Enhanced Imaging}.
The {\em NQIT} hub with its headquarter at Oxford University is developing a networked approach to quantum technologies as described above. Finally,  the {\em Quantum Communications} hub, led by York University, is developing quantum communication protocols for secure communication for communication with security based on quantum mechanical principles.

The European Commission has just announced an aspiration for a {flagship} research and development programme focussing on the  acceleration of quantum technologies research and development across Europe, and there are plans for similar large-scale research investments around the world. With financial support on this scale, the future for the development of these ambitious and transformative technologies seems bright.

\medskip
\emph{Acknowledgements.---}
We acknowledge financial support from the Royal Society, the UK EPSRC (EP/K034480/1 and EP/K004077/1), the Leverhulme Trust Research Grant (Project RPG-2014-055), European Research Council under the EU's Seventh Framework Programme (FP/2007-2013) / ERC Grant Agreement No. 308253, and the People Programme (Marie Curie Actions) of the EU's Seventh Framework Programme (FP7/2007-2013).

\end{document}